\documentclass[preprint,superscriptaddress,amsmath,showpacs,amssymb]{revtex4-1}
\usepackage[dvips,final]{graphicx}

\usepackage{physics,bm}
\usepackage{graphicx}

\usepackage{color}

\newcommand{\BM}{\begin{pmatrix}}
\newcommand{\EM}{\end{pmatrix}}

\bmdefine{\bx}{x}
\bmdefine{\by}{y}
\bmdefine{\bz}{z}
\bmdefine{\bl}{\lambda}
\bmdefine{\bn}{n}
\bmdefine{\bk}{k}

\newcommand{\ta}{\tilde{a}}
\newcommand{\txi}{\tilde{\xi}}

\newcommand{\bara}{\bar{a}}
\newcommand{\barxi}{\bar{\xi}}

\newcommand{\barS}{\bar{S}}

\newcommand{\barg}{\bar{g}}
\newcommand{\hH}{\hat{H}}
\newcommand{\dH}{\delta H}
\newcommand{\domega}{\delta \omega}
\newcommand{\mrmT}{\mathrm{T}}
\newcommand{\mrmH}{\mathrm{H}}

\newcommand{\mcalP}{\mathcal{P}}
\newcommand{\ppt}{\frac{\partial}{\partial t}}
\newcommand{\exv}[2]{\left\langle #1 \left| #2 \right| #1 \right\rangle}
\newcommand{\ints}{\int_{-\infty}^{t} \!\!\!\!\!\! ds}

\newcommand{\twomatrix}
  [4]{\mbox{$ {\displaystyle \left(
  \begin{matrix}{#1}&{#2}\cr{#3}&{#4}\end{matrix}
  \right) }$}}

\newcommand{\colmatrix}
  [2]{\mbox{$ {\displaystyle \left(
  \begin{matrix}{#1}\cr{#2}\end{matrix} \right) }$}}
\newcommand{\rowmatrix}
  [2]{\mbox{$ {\displaystyle \left(
  \begin{matrix}{#1}&{#2}\end{matrix}\right) }$}}

\newcommand{\threematrix}
  [9]{\mbox{$ {\displaystyle \left(
  \begin{matrix}{#1}&{#2}&{#3}\cr{#4}&{#5}&{#6}
   \cr{#7}&{#8}&{#9}\end{matrix}
  \right) }$}}

\newcommand{\threecolmatrix}
  [3]{\mbox{$ {\displaystyle \left(
  \begin{matrix}{#1}\cr{#2}\cr{#3}\end{matrix}\right) }$}}
\newcommand{\threerowmatrix}
  [3]{\mbox{$ {\displaystyle \left(
  \begin{matrix}{#1}&{#2}&{#3}\end{matrix}\right) }$}}

\begin{document}

\markboth{Kuwahara, Nakamura, Yamanaka}
{Self-Energy Renormalization for Inhomogeneous
Nonequilibrium Systems}

\title{SELF-ENERGY RENORMALIZATION FOR INHOMOGENEOUS
NONEQUILIBRIUM SYSTEMS\\
 AND FIELD EXPANSION VIA
COMPLETE SET OF TIME-DEPENDENT WAVE FUNCTIONS}

\author{Y.~Kuwahara}

\address{Department of Electronic and Physical Systems, Waseda
University\\
Tokyo 169-8555, Japan\\
a.kuwahara1224@asagi.waseda.jp}

\author{Y.~Nakamura\footnote{present address: 
Nagano Prefectural Kiso Seiho High School, Nagano 397-8571, Japan}}

\address{Institute of Condensed-Matter Science, Waseda
University\\
 Tokyo 169-8555, Japan\\ 
yusuke.n@asagi.waseda.jp}

\author{Y.~Yamanaka}

\address{Department of Electronic and Physical Systems, Waseda
University\\
Tokyo 169-8555, Japan\\
yamanaka@waseda.jp}

\begin{abstract}
The way to determine the renormalized energy of inhomogeneous systems
of a quantum field under an external potential
is established for both equilibrium and nonequilibrium scenarios 
based on Thermo Field Dynamics. The key step is to find an extension
of the on-shell concept valid in homogeneous case.
In the nonequilibrium case, we expand the field operator 
by time-dependent wave functions that are solutions of
the appropriately chosen differential equation, 
synchronizing with temporal change
of thermal situation, and 
the quantum transport equation is derived from the 
renormalization procedure. Through 
numerical calculations of a triple-well model with a reservoir, we show that the number distribution and the time-dependent wave functions
are  relaxed  consistently
to the correct equilibrium forms at the long-term limit.
\end{abstract}


\maketitle

\section{Introduction}

A sound formulation for nonequilibrium systems is desirable in quantum field theory, and will find many applications in wide areas of physics. As regards the testing of such theoretical formulations, systems of trapped, cold atomic gases are particularly ideal,\cite{PethickSmith,GriffinBook} as
their thermal processes move very slowly and can be observed 
experimentally.\cite{Davis}$^-$\cite{LesHouches} In addition, 
the experimental results 
can be compared with theoretical calculations.\cite{Jin}$^-$\cite{Bezett}
Another attractive feature is that various
nonequilibrium scenarios can be realized in experiments of cold atomic gases.

The two well-known real-time formalisms of 
 nonequilibrium quantum field system are Thermo Field Dynamics (TFD)\cite{UMT,AIP} and 
the closed time path (CTP) approach.\cite{Schwinger}$^-$\cite{Chou} 
Our arguments in this paper depend entirely on key
concepts such as the quasiparticle, the representation space 
(Fock space), and the renormalization, which are closely related to each other.
As TFD that is a canonical formalism 
is constructed on explicit uses of operator and representation (Fock) space,
the above key concepts must be addressed squarely and their implications 
in TFD have been argued and refined in various ways. On the other hand, CTP, mostly 
given in the path-integral, is 
formulated only in terms of Green's functions, and the roles
of operators and representation space are indirect. Thus, TFD is advantageous over CTP
for our purpose.

The renormalization performed for nonequilibrium 
TFD, in which every degree of freedom is doubled and the propagator and self-energy have thermal superscripts $(\mu,\nu=1,2)$, determines
the renormalized excitation energies
from their $(1,1)$- and/or $(2,2)$-components and derives the 
quantum transport equation from their $(1,2)$-component.
The quantum transport equation was first derived from the  
renormalized condition on the self-energy in the lowest-order perturbation
for the homogeneous system.\cite{AIP,YUNA} Then, this equation was generalized by Chu and Umezawa\cite{ChuUmezawa} as the diagonalization condition on the full propagator for homogeneous systems. 
Note that application of the
diagonalization condition to the full propagator at higher orders
is inconsistent with the equilibrium theory,\cite{AnnPhys331} 
because this application would imply that the Heisenberg 
and unperturbed number 
distributions are equal at the equilibrium limit. 
Thus, to maintain consistency with the equilibrium theory, 
we seek a new diagonalization condition on the on-shell
self-energy.
Extension of the method to systems of cold atomic gases that are inhomogeneous because
 of the trapping potentials is not straightforward, because the loss of translational symmetry obscures definitions of the 
on-shell self-energy on which the renormalization condition is to be imposed. 
More specifically, the self-energy in the $k_0$-space 
conjugate to time is, 
in general, non-diagonal matrices 
$\Sigma_{\ell_1\ell_2}(k_0)$ with indices $\ell$ 
of single quasiparticle states, and we are then uncertain as to what is 
the on-shell,
$k_0= \omega_{\ell_1}\,$, $\omega_{\ell_2}$ or something between them, which
is in contrast to homogeneous case in which $\ell$ is a momentum index ${\bk}$
and we have the diagonal $\Sigma_{\bk_1 \bk_2}(k_0)={\bar \Sigma}_{\bk_1}(k_0) 
\delta_{\bk_1\bk_2}$ owing to the momentum conservation law, and the on-shell
is unambiguously achieved by 
putting $k_0=\omega_{\bk_1}$.
We have been attempting several 
formulations,\cite{AnnPhys331,AnnPhys325}$^-$\cite{APPCNakamura}
but they are not entirely satisfactory. 

In the meantime, understanding of the nonequilibrium TFD formulation
was expanded.\cite{AnnPhys331,PhysLetA377} 
By deriving the nonequilibrium TFD formulation from the 
superoperator formalism presented in Ref.~\cite{AnnPhys331}, 
we clarified that this formulation is based on the existence of 
a quasiparticle 
picture at each instant of time. In addition, a thermal causality holds, 
where the macroscopic quantities such as the number distribution should 
affect the microscopic motions in the future only. 
The time-dependent unperturbed 
representation in the interaction picture
of nonequilibrium TFD, which is self-consistently selected by 
the renormalization condition, provides the optimum approximation 
method. In addition, we demonstrated in Ref.~\cite{PhysLetA377}
that the thermal causality is an important requirement in 
deriving the nonequilibrium TFD directly from the classical Hamilton
principle for nonconservative systems with doubled degrees of
freedom.\cite{Galley} The thermal causality was also helpful 
as a guiding principle when we devised a new definition of the on-shell 
self-energy in Ref.~\cite{AnnPhys331}.

The two main problems of nonequilibrium TFD 
to be considered in the present study are (i) the renormalization condition on the 
self-energy and 
(ii) the appropriate choice of 
a complete set of time-dependent eigenfunctions. We 
present consistent solutions to both the problems.
As regards problem (i), we consider the renormalization condition 
for both equilibrium and nonequilibrium inhomogeneous systems, noting that 
the nonequilibrium formulation should approach 
the equilibrium formulation in the long-term limit. The derived 
quantum transport equations describe the relaxation of the number
distribution to the equilibrium form. The problem (ii) arises from
the consideration that as the quasiparticle picture should change in time 
during nonequilibrium processes, particularly when a 
time-dependent condensate is present, and the field operator should be
expanded in an appropriate complete set of time-dependent 
eigenfunctions.\cite{Matsumoto2001,AnnPhys325} Note that the question as to 
whether the time-dependent eigenfunctions converge to 
the stationary eigenfunctions at equilibrium, following 
the given differential equation, is not trivial at all. 
The time-dependent eigenequation is coupled to both of 
the quantum transport equation
and the equation to determine the renormalized energy. In this study, 
we establish the time-dependent eigenequation that reduces to 
the usual stationary
 eigenequation at equilibrium. To confirm the good performance of our formulation, we take a rather simple model of a triple-well with reservoir and perform numerical calculations. 
 
The remainder of this paper is organized as follows. A brief review of the renormalization
condition of the stationary homogeneous system is given 
in Section~\ref{sec-Homogeneous}.
In Section~\ref{sec-InhEqui}, we present the 
renormalization condition of the inhomogeneous
system in equilibrium, so as to determine the renormalized excitation energy
using the TFD formalism. The main part of this paper is 
Section~\ref{sec-InhNon},
in which we consider the nonequilibrium inhomogeneous system, 
establishing the equations for  the time-dependent eigenfunctions
and giving the renormalization conditions on the $(1,2)$-component 
and the $(1,1)$- and/or $(2,2)$-components of the on-shell self-energy; hence, the quantum transport equation is derived and the renormalized
time-dependent excitation energy is fixed. The formulation obtained in Section~\ref{sec-InhNon} is applied to a triple-well model with reservoir in 
Section~\ref{sec-model}.
Some analytic expressions of the model are derived in Appendix.
 Numerical calculations of the model are performed and 
the results are presented in Section~\ref{sec-model}. 
Section~\ref{sec-Summary} is devoted to a summary.

\section{Renormalization condition of stationary homogeneous system}
\label{sec-Homogeneous}

For later comparison, we first outline the well-known renormalization condition
of the stationary homogeneous system to 
determine the energy counter term in the Hamiltonian at zero temperature. 

We consider a homogeneous system comprised of a bosonic quantum field $\psi(x)$ having the Hamiltonian 
\begin{align}
	&H^h = H_0^h + H_{\rm int} \label{eq:HamiltonianUniform}\,, \\
	& H_0^h=\int\! d^3x\, \psi^{\dagger}(x) 
	\left( - \frac{\bm{\nabla}^2}{2m}-\mu\right) \psi(x)\,, \label{eq:H0h}
\end{align}
where $x = (\bx ,t)$ and $m$ and $\mu$ represent the mass 
and chemical potential, respectively.
The interaction Hamiltonian nonlinear in $\psi(x)$ is not specified, because 
its explicit form is not essential to our discussion.  We set $\hbar = 1$ throughout this paper.
The canonical commutation relations are 
\begin{align}
	&[\psi(x), \psi^{\dagger}(x')]|_{t=t'} = \delta(\bx - \bx') \,,
	 \label{eq:CCR1}\\
	&[\psi(x), \psi(x')]|_{t=t'} =
	 [\psi^{\dagger}(x),\psi^{\dagger}(x')]|_{t=t'} = 0 \,.
	\label{eq:CCR2}
\end{align}
It is customary to expand $\psi(x)$ in momentum eigenfunctions, such that
\begin{equation}
\psi(x) = \frac{1}{(2\pi)^{3/2}}\int\! d^3k\,
 e^{i\bk \cdot \bx} a_{\bk}(t) \,,
\end{equation}
because the translational symmetry of the homogeneous system implies
conservation of the total momentum and the momentum is a good quantum number.
Then, $H_0^h$ is diagonalized as
\begin{align}
	H_0^h = \int\! d^3 k\, \omega_{k}^0 a_{\bk}^{\dagger} a_{\bk} \,,
\end{align}
with the unperturbed bare energy $\omega_{k}^0 = \frac{k^2}{2m}-\mu$.
The interaction shifts the bare energy into a renormalized energy,
denoted by $\omega_{\bk}$,
which is observed. Because of the momentum conservation, 
both the renormalized unperturbed Hamiltonian $H_u^h$ and 
the energy counter term $\dH^h$
are diagonal with respect to the momentum index, such that
\begin{align}
	& H_u^h = H_0^h + \dH^h=\int\! d^3 k\,
	 \omega_{\bk} a_{\bk}^{\dagger} a_{\bk}\,, \\
	&\dH^h = \int\! d^3 k\, \domega_{\bk} a_{\bk}^{\dagger} a_{\bk} \,.
\end{align}
Note that the interaction Hamiltonian is not $H_{\rm int}$, 
but $H^h_I$, where
\begin{equation}
	H^h_I = H_{\rm int} - \dH^h	\,.
\end{equation}

The renormalization, or the determination of the counter term,
 is performed as follows. 
We take the Fourier transform of the self-energy $\Sigma$,
 which depends on the relative time and coordinate only, with respect to the 
relative time but with respect to the  the two coordinates separately, such that
\begin{align}
&\Sigma ( \bx-\bx',t-t') = \int\!\frac{dk_0 d^3kd^3k'}{(2\pi)^7}\,
e^{-ik_0(t-t')+ i(\bx\cdot \bk-\bx'\cdot \bk')} \, 
{\bar \Sigma}_{\bk \bk'}(k_0) \,, \\
&{\bar \Sigma}_{\bk \bk'}(k_0)={\bar \Sigma}_{\bk}(k_0)\delta(\bk-\bk')\,,
\end{align}
which is the sum of the loop and counter term contributions,
\begin{align}
 {\bar \Sigma}_{\bk}(k_0)={\bar \Sigma}_{\bm k}^{\rm loop} (k_0)
+ {\bar \Sigma}_{\bm k}^{-\delta H}, \qquad \mbox{with}\qquad
{\bar \Sigma}_{\bk}^{-\delta H} = -\delta \omega_{\bk}\,.
\end{align}
We can determine $\delta \omega_\bk$ consistently, by imposing 
the on-shell renormalization condition
\begin{align}
{\bar \Sigma}_{\bk}(k_0=\omega_\bk)=0 \,.
\end{align}

We also review the renormalization of the homogeneous system in equilibrium, 
using TFD. In TFD, every degree of freedom is doubled and the thermal 
Bogoliubov transformation is introduced, such that 
\begin{equation}
	a_{\bk}^{\mu}
	= B^{-1, \mu \nu}[n_{\bk}] \xi_{\bk}^{\nu} \,,\hspace{10pt}
	\bara_{\bk}^{\nu}
	=  \barxi_{\bk}^{\mu} B^{\mu\nu}[n_{\bk}] \,,
\end{equation}
where the thermal doublet notations are used:
\begin{align}
	&a_{\bk}^{\mu} = {\colmatrix {a_{\bk}} 
	{\ta_{\bk}^{\dagger}}}^{\mu}\,,\hspace{10pt}
	\bara_{\bk}^{\nu} =
	 {\rowmatrix {a_{\bk}^{\dagger}}{ -\ta_{\bk}}}^{\nu} \,, \\
	&\xi_{\bk}^{\mu} = {\colmatrix{\xi_{\bk}}{\txi_{\bk}^{\dagger}}}^{\mu}
	\,,\hspace{10pt}
	\barxi_{\bk}^{\nu} = {\rowmatrix{\xi_{\bk}^{\dagger}}
	{ -\txi_{\bk}}}^{\nu} \,, \\
	&B^{\mu  \nu}[n_{\bk}]= {\twomatrix{1+n_{\bk}}{ -n_{\bk}} 
	{-1}{1}}^{\mu \nu}\,, 
	\hspace{10pt}
	B^{-1, \mu \nu}[n_{\bk}] = {\twomatrix{1}{n_{\bk}}{1}
	{1+n_{\bk}}}^{\mu \nu}\,.
\end{align}
The dummy thermal superscripts imply the Einstein summation convention.
The matrix $B^{\mu\nu}$ is called the ``thermal Bogoliubov matrix".
The choice of the above form of $B^{\mu\nu}$ corresponds to the
$\alpha=1$ representation of TFD\cite{AIP}.
The canonical commutation relations are
\begin{align}
[ a_\bk^\mu \,,\, {\bar a}_{\bk'}^\nu] = 
[ \xi_\bk^\mu \,,\, {\bar \xi}_{\bk'}^\nu]=\delta_{\mu\nu}\delta_{\bk\bk'}
\, .
\end{align}
Further, the total Hamiltonian for time translation 
of both the non-tilde and tilde operators is 
\begin{align}
\hH = H - \tilde{H}\,,
\label{eq:hatH}
\end{align}
which is called the ``hat total Hamiltonian". Here, $\tilde{H}$ is obtained by replacing the non-tilde operators and
the c-number coefficients with corresponding tilde operators and  
their complex conjugates, respectively.

 The $\xi$-operators annihilate the thermal vacuum $\ket{0}$, where
\begin{align}
\xi_{\bk} \ket{0} = \txi_{\bk} \ket{0} = 0 \,,
\qquad \bra{0}\xi^\dagger_{\bk}  = \bra{0}\txi^\dagger_{\bk}  = 0
\,.
\end{align}
The thermal averages are given by the pure state averages of the thermal vacua. In particular, the number density $n_{\bk}$ is
\begin{equation}
	n_{\bk} = \exv{0}{ a_{\bk}^{\dagger} a_{\bk}} \,,
\end{equation}
which is the Bose--Einstein distribution in equilibrium.

The renormalization condition is applied to the self-energy 
of the $\xi$-operators rather than that of the $a$-operators, 
because the $\xi$-operators represent the 
quasiparticles in thermal scenarios. 
The full and unperturbed
propagators of the $\xi$-operators, denoted by $g^{\mu\nu}_{\bk}$ and
 $d^{\mu\nu}_{\bk}$, respectively, are defined as 
\begin{align}
	&g_{\bk_1}^{\mu \nu}(t_1 - t_2) \delta (\bk_1 - \bk_2)
	= -i \exv{0}{\mrmT \left[ \xi_{\mrmH \bk_1}^{\mu}(t_1)  
	\barxi_{\mrmH \bk_2}^{\nu}(t_2) \right]}\,, \\
	&d_{\bk_1}^{\mu \nu}(t_1 - t_2) \delta (\bk_1 - \bk_2)
	= -i \exv{0}{\mrmT \left[\xi_{\bk_1}^{\mu}(t_1)  
	\barxi_{\bk_2}^{\nu}(t_2) \right]}\,,
\end{align}
where $\mrmT$ represents a time-ordered product and 
the suffix $\mrmH$ implies that the operator
is that of the Heisenberg picture. The self-energy 
$S_{\bk}^{\mu \nu}(t_1 - t_2)$ is
defined through the Dyson equation, as
\begin{equation}
	g_{\bk}^{\mu \nu}(t_1 - t_2)
	= d_{\bk}^{\mu \nu}(t_1 - t_2) + 
	\int\! ds_1 ds_2\, d_{\bk}^{\mu \mu'}(t_1 - s_1) 
	S_{\bk}^{\mu' \nu'}(s_1 - s_2) g_{\bk}^{\nu' \nu}(s_2 - t_2) \,.
\end{equation}
Taking the Fourier transform of $S_{\bk}^{\mu \nu}(t_1 - t_2)$ 
with respect to the relative time $\tau = t_1 - t_2$, we obtain 
\begin{equation}
	\barS_{\bk}^{\mu \nu}(k_0) = \int d\tau S_{\bk}^{\mu \nu}(\tau)
	 e^{ik_0 \tau}\,.
\end{equation}
We apply the on-shell renormalization condition 
to the real part of the $(1,1)$-component, such that 
\begin{equation}
\label{eq:RenormCondUniform}
	{\rm Re} \left[\barS_{\bk}^{11}(\omega_\bk)\right] = 0 \,.
\end{equation}
Here, the counter term contribution is $\barS_{\bk}^{\mu\nu, -\dH}(\omega_\bk)
=-\delta_{\mu \nu} \domega_{\bk} $, and $\domega_{\bk}$ is determined.
As $\barS_{\bk}^{11}(\omega_\bk)
= \barS_{\bk}^{22, \ast}(\omega_\bk)$,
the condition ${\rm Re} [\barS_{\bk}^{22}(\omega_\bk)]= 0$ 
adds no constraint. The on-shell self-energy $\barS_{\bk}^{11}(\omega_\bk)$
is complex in general, but we do not renormalize its imaginary component, 
because no sound prescription or renormalizing imaginary component of the energy is known.

\section{Renormalization condition of inhomogeneous system in equilibrium}
\label{sec-InhEqui}

In this section, we consider an equilibrium system that is inhomogeneous
as a result of an external potential $V(\bx)$, along with its renormalization condition, so as to 
determine the energy counter term. This renormalization condition is
extended to the nonequilibrium case in the next section.

In this case, the free Hamiltonian is given by
\begin{align}
	H_0 = \int\! d^3x\,
	\psi^{\dagger}(x) h_0(\bx) \psi(x), \qquad \mbox{with}
	\qquad h_0(\bx)= - \frac{\bm{\nabla}^2}{2m}+ V(\bx)-\mu\,.
\end{align}
Contrary to the homogeneous case, the expansion of $\psi(x)$ into the momentum 
eigenfunctions is not useful. We may expand $\psi(x)$ in a complete set of
the eigenfunctions for $h_0(\bx)$ so that $H_0$ is diagonalized, but 
the renormalized unperturbed Hamiltonian for the inhomogeneous system
with the counter term $\delta H$, the general form of which is 
\begin{align}
	& H_u =H_0+ \dH \,, \\
	& \dH(t) = \int\! d^3x d^3x' \, 
	\psi^{\dagger}(\bx,t) \domega(\bx, \bx') \psi(\bx',t) \,,
\end{align}
is non-diagonal. The diagonal form of $H_u$ is essential, but
the diagonal forms of $H_0$ and $\delta H$ are not required. Therefore, we adopt the complete orthonormal set 
of eigenfunctions $\left\{ u_{\ell}(\bx) \right\}$, where
\begin{align}
	\int\! d^3x'\, h_u(\bx , \bx') u_{\ell}(\bx') 
	= \omega_{\ell} u_{\ell}(\bx) \,,
	\qquad 
	h_u(\bx, \bx') = \delta(\bx - \bx')h_0(\bx) + \domega(\bx, \bx') \,,
	\label{eq:hu}
\end{align}
with $\domega(\bx, \bx') =\domega^\ast(\bx', \bx)$, so as to expand $\psi(x)$ as
\begin{equation}
	\psi(x) = \sum_{\ell} u_{\ell}(\bx) a_{\ell}(t)	 \,.
\end{equation}
In fact, $H_u$ has a diagonal form, where
\begin{align}
	H_u = \sum_{\ell} \omega_{\ell} a_{\ell}^{\dagger} a_{\ell} \,.
\label{eq:Hu}
\end{align}
The counter term is written as
\begin{align}
	&\dH(t) = \sum_{\ell_1 \ell_2} \domega_{\ell_1 \ell_2}
	 a_{\ell_1}^{\dagger}(t) a_{\ell_2}(t) \,, \\
	& {\rm with} \quad 
	\domega_{\ell_1 \ell_2} = \int\! d^3x d^3 x'\,
	 u_{\ell_1}^\ast(\bx) \domega(\bx, \bx') u_{\ell_2}(\bx') \,.
\end{align}

We provide a TFD formulation of the inhomogeneous system above 
in thermal scenarios (both equilibrium and nonequilibrium), 
doubling every degree 
of freedom and introducing the thermal Bogoliubov transformation 
with the number distribution $n_\ell$ and the thermal
vacuum, in parallel with that for the homogeneous system 
given in the previous section.
For the equilibrium case, the parameter $n_\ell$ is the stationary Bose-Einstein distribution, i.e.,
\begin{align}
n_\ell= \frac{1}{e^{\beta \omega_\ell}-1} \,.
\end{align}	
However, for the nonequilibrium case, this parameter is a unknown function of $t$, denoted by $n_\ell(t)$. The full and unperturbed propagators 
of the $\xi$-operators are, in general, defined as
\begin{align}
	&g_{\ell_1 \ell_2}^{\mu \nu}(t_1 , t_2)
	= -i \exv{0}{ \mrmT \left[ \xi_{\mrmH \ell_1}^{\mu}(t_1) 
	\barxi_{\mrmH \ell_2}^{\nu}(t_2) \right]} \,, \\
	&d_{\ell_1 \ell_2}^{\mu \nu}(t_1,t_2)
	= -i \exv{0}{ \mrmT \left[ \xi_{\ell_1}^{\mu}(t_1) 
	\barxi_{\ell_2}^{\nu}(t_2) \right]} \,,
\end{align}
respectively. The self-energy $S_{\ell_1 \ell_2}^{\mu \nu}(t_1 , t_2)$
is defined in the  Dyson equation,
\begin{align}
	&g_{\ell_1 \ell_2}^{\mu \nu}(t_1 , t_2)
	= d_{\ell_1 \ell_2}^{\mu \nu}(t_1 , t_2)  \notag \\
 	&\qquad +\sum_{m_1 m_2} \int\! ds_1 ds_2\, 
	d_{\ell_1 m_1}^{\mu \mu'}(t_1 , s_1) 
	S_{m_1 m_2}^{\mu' \nu'}(s_1 , s_2) g_{m_2 \ell_2}^{\nu' \nu}
	(s_2 , t_2) \,,
\end{align}
and has the following properties:
\begin{align}
	&S_{\ell_1\ell_2}^{\mu\nu}(t_1,t_2)=
	\twomatrix {S^{11}_{\ell_1\ell_2}(t_1, t_2)}
	{S^{12}_{\ell_1\ell_2}(t_1,t_2)}
	{0}{S^{22}_{\ell_1\ell_2}(t_1,t_2)} \,, \\
	& S^{11}_{\ell_1\ell_2}(t_1,t_2)\, \propto 
	\, \theta(t_1-t_2) \,, \qquad
	S^{22}_{\ell_1\ell_2}(t_1,t_2)\, \propto 
	\, \theta(t_2-t_1) \,, \\
	&S^{11}_{\ell_1\ell_2}(t_1,t_2)=S^{22, \ast}_{\ell_2\ell_1}
	(t_2,t_1)\,, \qquad S^{12}_{\ell_1\ell_2}(t_1,t_2)
	=-S^{12, \ast}_{\ell_2\ell_1}(t_2
	, t_1)\,.
\end{align}
Note that $S^{21}_{\ell_1\ell_2}=0$, which is inherent to 
the $\alpha=1$ representation\cite{AIP}. Further, the phase symmetry is assumed to be unbroken.

In this section, we restrict ourselves to equilibrium cases in which 
the propagators and self-energy are functions of the 
relative time $\tau=t_1-t_2$.
Then, the Fourier transformation
of $S_{\ell_1 \ell_2}^{\mu \nu}(\tau)$ with respect to $\tau$ is defined by
\begin{align}
	\barS_{\ell_1 \ell_2}^{\mu \nu}(k_0) =
	 \int\! d\tau\,  S_{\ell_1 \ell_2}^{\mu \nu}(\tau) e^{i k_0 \tau} \,,
\end{align}
which is the sum of the loop contribution (denoted by
$\barS_{\ell_1 \ell_2}^{\mu \nu, {\rm loop}}(k_0)$) and 
the contribution from $-\dH$, where
\begin{equation}
	\barS_{\ell_1 \ell_2}^{\mu \nu, -\dH}(k_0) 
	= - \domega_{\ell_1 \ell_2}
	{\twomatrix {1}{ n_{\ell_2} - n_{\ell_1}}{ 0}{1}}^{\mu \nu} \,.
	\label{eq:barSdeltaH}
\end{equation}
The loop contribution is expressed in spectral form \cite{AIP,AnnPhys326},
 such that
\begin{align}
	&\barS_{\ell_1 \ell_2}^{11, {\rm loop}}(k_0)=
	{\bar \sigma}_{\ell_1\ell_2}(k_0)-
	i \pi \sigma_{\ell_1\ell_2}(k_0) \,, \\
	&\barS_{\ell_1 \ell_2}^{22, {\rm loop}}(k_0)=
	{\bar \sigma}_{\ell_1\ell_2}(k_0)
	+i \pi \sigma_{\ell_1\ell_2}(k_0) \,, \\
	&\barS_{\ell_1 \ell_2}^{12, {\rm loop}}(k_0)=
	(n_{\ell_2}-n_{\ell_1}) 
	{\bar \sigma}_{\ell_1\ell_2}(k_0)
	-i \pi\{ n_{\ell_1}+n_{\ell_2}-2n(k_0)\} 
	\sigma_{\ell_1\ell_2}(k_0) \,,
\end{align} 
where $\sigma_{\ell_1\ell_2}(\kappa)$ is the spectral function 
with the Hermitian  property $\sigma_{\ell_1\ell_2}(\kappa)= 
\sigma_{\ell_2\ell_1}^\ast(\kappa)$, and
\begin{align}
	{\bar \sigma}_{\ell_1\ell_2}(k_0)=\int_{-\infty}^\infty
	d\kappa\, {\mathcal P}\frac {\sigma_{\ell_1\ell_2}(\kappa)}
	{k_0-\kappa}\, .
\end{align}

No definite renormalization condition for the inhomogeneous system in
thermal scenarios
(both equilibrium and nonequilibrium) has been established.
The difficulty lies in the fact that the self-energy and the counter term 
are first non-diagonal in the index of the quantum number $\ell$.
Our previous attempt was to apply the ``on-shell" 
renormalization condition only 
to the elements, which are diagonal in both the index $\ell$ and the thermal index, i.e.,
${\rm Re}{\bar S}^{11}_{\ell\ell}(\omega_\ell)=
 {\rm Re}{\bar S}^{22}_{\ell\ell}(\omega_\ell)=0$, following the 
homogeneous case.  However, it must be mentioned that 
$\delta \omega_{\ell_1\ell_2}$ $(\ell_1 \neq \ell_2)$ 
are then left undetermined. 
Furthermore, the element  
${\bar S}^{12}_{\ell_1\ell_2}(\omega_\ell)$ is 
non-vanishing in general. This is undesirable, as the condition 
$S^{12}[\omega_\ell:t] =0$ (this notation will be defined 
in Eq.~(\ref{eq:omegaRep}))
is used to derive the quantum transport equation in the nonequilibrium case,
and the equilibrium theory should be a stationary limit of
the nonequilibrium case.

We here propose a new renormalization condition for an inhomogeneous equilibrium
system that determines all the elements of $\delta \omega_{\ell_1\ell_2}$,
and that simultaneously induces 
vanishing of the on-shell ${\bar S}^{12}_{\ell_1\ell_2}$.
The concept of the on-shell energy becomes vague in an inhomogeneous system,
because the self-energy ${\bar S}^{12}_{\ell_1\ell_2}(k_0)$ is not diagonal
with respect to the index $\ell$ and no well-grounded definition of the on-shell
($k_0= \omega_{\ell_1}\,,\,\omega_{\ell_1}$\, or something between them) is known.
In order to define the on-shell energy in the present case, 
we recall that the full propagator of the $\alpha=1$ representation is
rewritten in the interaction picture as
\begin{align}
	g_{\ell_1\ell_2}^{\mu\nu}(t_1,t_2)&=-i\theta(t_1-t_2)\bra{0}
	 \xi_{\ell_1}^\mu(t_1){\hat U}(t_1,t_2)
	 {\bar \xi}_{\ell_2}^\nu(t_2){\hat U}(t_2,-\infty)\ket{0} \notag \\
	&\quad -i\theta(t_2-t_1)\bra{0}{\bar \xi}_{\ell_2}^\nu(t_2)
	 {\hat U}(t_2,t_1) \xi_{\ell_1}^\mu(t_1){\hat U}(t_1,-\infty)
	\ket{0}\,, 
\label{eq:fullg}
\end{align}
because $\bra{0} {\hat U}(\infty,t)=\bra{0}$\cite{AIP}. 
Here ${\hat U}(t,t')$ is the time-translation operator  and is 
given explicitly by
\begin{align}
{\hat U}(t,t')= \mrmT \left[ \exp 
\left[-i \int_{t'}^tds\, {\hat H}_I(s) \right] \right]\,,
\end{align} 
with the hat interaction Hamiltonian 
${\hat H}_I(t)={\hat H}(t)-{\hat H}_u(t)$. 
Its form in nonequilibrium case is seen in Eq.~(\ref{eq:hatHint}) below.
Note that $\bra{0} {\hat H}_I(t)$ =0 in the $\alpha=1$ representation
of TFD\cite{AIP}. Equation (\ref{eq:fullg}) indicates that the retarded and advanced parts of the full propagator 
exclusively describe the transitions into the fixed final states of the 
quasiparticle $\bra{0}\xi_{\ell_1}^\mu$ with $\mu=1$,
and $\bra{0}{\bar \xi}_{\ell_2}^\nu$ with $\nu=2$, respectively. 
Therefore, we define the on-shell self-energy in the stationary case, setting
 $k_0=\omega_{\ell_1}$ and $\omega_{\ell_2}$
for its retarded and advanced parts, respectively. Explicitly, the on-shell self-energies in equilibrium are
${\bar S}_{\ell_1\ell_2}^{11}(\omega_{\ell_1})$, 
${\bar S}_{\ell_1\ell_2}^{22}(\omega_{\ell_2})$, 
${\bar S}_{\ell_1\ell_2}^{12,+}(\omega_{\ell_1})$, and 
${\bar S}_{\ell_1\ell_2}^{12,-}(\omega_{\ell_2})$, 
where
\begin{align}
{\bar S}_{\ell_1\ell_2}^{12}(k_0)
={\bar S}_{\ell_1\ell_2}^{12,+}(k_0)
+{\bar S}_{\ell_1\ell_2}^{12,-}(k_0) \,,
\end{align}
and ${\bar S}_{\ell_1\ell_2}^{12,+}(k_0)$ and 
${\bar S}_{\ell_1\ell_2}^{12,-}(k_0)$ are retarded and advanced 
parts, respectively. 
In this scenario, the renormalization conditions such as 
${\bar S}_{\ell_1\ell_2}^{11}(\omega_{\ell_1})=0$ or
${\bar S}_{\ell_1\ell_2}^{22}(\omega_{\ell_2})=0$ would be inconsistent,
as ${\bar S}_{\ell_1\ell_2}^{11,{\rm loop}}(\omega_{\ell_1})$ and 
${\bar S}_{\ell_1\ell_2}^{22,{\rm loop}}(\omega_{\ell_2})$ are non-Hermitian
matrices and cannot cancel the Hermitian counter term.
Thus, we devise a renormalization condition on 
the following combination of on-shell self-energies:
\begin{align}
	0&={\bar S}_{\ell_1\ell_2}^{11}(\omega_{\ell_1})
	+ {\bar S}_{\ell_1\ell_2}^{22}(\omega_{\ell_2}), \notag \\
	&= -2\delta \omega_{\ell_1\ell_2}
	+{\bar \sigma}_{\ell_1\ell_2}(\omega_{\ell_1})
	+{\bar \sigma}_{\ell_1\ell_2}(\omega_{\ell_2})
	-i\pi \sigma_{\ell_1\ell_2}(\omega_{\ell_1})
	+i\pi \sigma_{\ell_1\ell_2}(\omega_{\ell_2}) \,.
	\label{eq:barS1122onshell}
\end{align}
As this matrix equation is Hermitian,  
all the elements $\delta \omega_{\ell_1\ell_2}$ are obtained 
consistently. Equation~(\ref{eq:barS1122onshell})
implies that no imaginary part of the diagonalized energy
is renormalized.
Taking the combination, or the Hermitian part of 
the on-shell ${\bar S}^{\mu\mu}_{\ell_1\ell_2}$,
corresponds to taking the real part of the on-shell self-energy
of the homogeneous system as in (\ref{eq:RenormCondUniform}), which
is rewritten equivalently as $\barS_{\bk}^{11}(\omega_\bk)
+\barS_{\bk}^{22}(\omega_\bk)=0$.

Next, we consider the on-shell self-energy 
with superscript $(1,2)$, i.e.,
${\bar S}_{\ell_1\ell_2}^{12,+}(\omega_{\ell_1})
+ {\bar S}_{\ell_1\ell_2}^{12,-}(\omega_{\ell_2})$,
which can be manipulated as
\begin{align}
	{\bar S}_{\ell_1\ell_2}^{12,+}(\omega_{\ell_1})
	+ {\bar S}_{\ell_1\ell_2}^{12,-}(\omega_{\ell_2})=
	(n_{\ell_2}-n_{\ell_1})
	\left\{{\bar S}_{\ell_1\ell_2}^{11}(\omega_{\ell_1})
	+ {\bar S}_{\ell_1\ell_2}^{22}(\omega_{\ell_2})\right\}
	=0\,,  	\label{eq:barS12onshell}
\end{align}
from
\begin{align}
	&{\bar S}_{\ell_1\ell_2}^{12,+}(\omega_{\ell_1})
	=(n_{\ell_2}-n_{\ell_1})\left\{ -\delta \omega_{\ell_1\ell_2}
	+{\bar \sigma}_{\ell_1\ell_2}(\omega_{\ell_1})
	-i\pi \sigma_{\ell_1\ell_2}(\omega_{\ell_1})\right\}\, ,
	\\
	&{\bar S}_{\ell_1\ell_2}^{12,-}(\omega_{\ell_2})=
	(n_{\ell_2}-n_{\ell_1})\left\{ -\delta \omega_{\ell_1\ell_2}
	+{\bar \sigma}_{\ell_1\ell_2}(\omega_{\ell_2})
	+i\pi \sigma_{\ell_1\ell_2}(\omega_{\ell_2})\right\}\, .
\end{align}
Thus, the renormalization condition (\ref{eq:barS1122onshell})
yields vanishing on-shell self-energy 
with superscript $(1,2)$
simultaneously.

This way we have obtained the one-shell renormalization 
conditions Eqs.~(\ref{eq:barS1122onshell}) and (\ref{eq:barS12onshell})
successfully in a sense that all the matrix elements (both $(\mu,\nu)$
and $(\ell_1,\ell_2)$ simultaneously) of the on-shell self-energy vanish.

\section{Renormalization condition of inhomogeneous system in nonequilibrium}
\label{sec-InhNon}

In this section, the renormalization method to determine 
the counter term for inhomogeneous systems in equilibrium is extended to
nonequilibrium scenarios. Now, the counter term
depends on time, as $\delta \omega (\bx,\bx',t)$. 
In addition, $h_u$ in Eq.~(\ref{eq:hu}) also depends on time, i.e.,
\begin{equation}
	h_u(\bx, \bx',t) = \delta(\bx - \bx')h_0(\bx)
	 + \domega(\bx, \bx',t) \,. \label{eq:hut}
\end{equation}
We begin with the equation for the unperturbed field
\begin{align}
	i \ppt \psi(x) = \int\!d^3 x'\,  h_u(\bx,\bx',t) \psi(x')
	|_{t= t'}	\,, 
\end{align}
and the equation for the quasiparticle
\begin{align}
	 i \frac{d}{dt}  a_\ell(t) = \omega_\ell(t) a_\ell(t) \,,
\end{align}
both of which are expected to be
generated by the unperturbed Hamiltonian
\begin{align}
	H_u &= \int\! d^3x d^3x' \, 
	\psi^{\dagger}(x) h_u(\bx,\bx',t) \psi(x')|_{t= t'}, \\
	&= \sum_{\ell} \omega_\ell(t) a_\ell^\dagger(t)a_\ell(t)\,.
\end{align}
Then, $\psi(x)$ is expanded in terms of the
time-dependent functions\cite{Matsumoto2001} $\{v_\ell(x)\}$, such that
\begin{align}
	\psi(x) = \sum_{\ell} v_{\ell}(x) a_{\ell}(t) \,,
	\label{eq:psiExpv}
\end{align}
with $v_\ell(t)$ satisfying
\begin{align}
	\label{eq:TDCS}
	&i \ppt v_{\ell}(x) = \int\!d^3 x'\,  h_u(\bx,\bx',t)
	 v_{\ell}(x')|_{t= t'}
	 - \omega_{\ell}(t) v_{\ell}(x) \,, \\
	&{\rm with} \quad
	\omega_{\ell}(t) = \int\! d^3 x d^3 x'\, 
	v_{\ell}^\ast(x) h_u(\bx, \bx',t)  v_{\ell}(x')|_{t= t'} \,.
	\label{eq:omegaellt}
\end{align}
This corresponds to Eq.~(\ref{eq:hu}) at the stationary limit.
It follows from Eq.~(\ref{eq:TDCS}) and the assumption of Hermiticity of
$\delta \omega(\bx,\bx',t)$ that
\begin{align}
	&\frac{d}{dt}
	\int\!d^3 x\, v_{\ell}^\ast(x) v_{\ell'}(x) =0 \,,
	\label{eq:vtOrtht}\\
	&\ppt \left[\sum_{\ell} v_{\ell}(\bx,t) v_{\ell}^\ast(\bx',t) 
	\right]=0\,. \label{eq:vtCompletet}
\end{align}
Consequently, $\{v_\ell(x)\}$ remains a complete orthonormal set
 at any time $t$, if it is taken to be a complete orthonormal set
at the initial time. This is necessary for consistency of the above
 formulation, given in Eqs.~(\ref{eq:hut}) -- (\ref{eq:omegaellt}).

We now turn to the nonequilibrium TFD
formalism by doubling every degree of freedom and introducing
the unknown time-dependent number distribution $n_\ell(t)$.
As $\delta \omega_{\ell_1\ell_2}(t)$ is determined by 
the renormalization condition,
the quantum transport equation for $n_\ell(t)$
is similarly derived from the renormalization condition in TFD. 

The hat unperturbed Hamiltonian 
of the system under consideration is 
\begin{align}
	\hH_u(t) = \hH_0(t) + \delta \hH(t) - \hat{Q}(t) \,,
	\label{eq:hatHut}
\end{align}
where $\hH_0(t)= H_0(t) -{\tilde H}_0(t)$\,,
$\delta \hH(t)=\delta H(t)-\delta {\tilde H}(t)$, and the thermal counter
term, which mixes non-tilde and tilde operators and drives the nonequilibrium thermal
changes, is
\begin{equation}
	\hat{Q}(t) = -i \sum_{\ell} \dot{n}_{\ell}(t) 
	\xi_{\ell}^{\dagger}(t) \txi_{\ell}^{\dagger}(t)\,.
	\label{eq:hatQ}
\end{equation}
As the hat total Hmiltonian is given in Eq.~(\ref{eq:hatH}), 
the hat interaction Hamiltonian is
\begin{align}
{\hat H}_I(t) ={\hat H}_{\rm int}(t)-\delta \hH(t)+\hat{Q}(t) \,.
\label{eq:hatHint}
\end{align}

The on-shell self-energy in equilibrium,
 on which the renormalization condition is imposed,
can be defined with regard to the Fourier component with respect to the relative time. However, the same approach cannot be employed in the nonequilibrium case, where the time-translation symmetry is lost. Instead, we consider the
following quantity, which is a functional\cite{AnnPhys331} of an arbitrary 
function  $\omega(t)$ as well as a function of $t$: 
\begin{align}
	\label{eq:omegaRep}
	&\barS_{\ell_1 \ell_2}^{\mu \nu}[\omega;t]
	=\barS_{\ell_1 \ell_2}^{\mu \nu,+}[\omega;t]
	+\barS_{\ell_1 \ell_2}^{\mu \nu,-}[\omega;t] \,,\\
	&\barS_{\ell_1 \ell_2}^{\mu \nu,+}[\omega;t]
	= \int \! d\tau \, \theta(\tau)
	 S_{\ell_1 \ell_2}^{\mu \nu}(t, t-\tau) 	
	e^{i \int_{t - \tau}^{t} \!\!\! ds\, \omega(s)} \,, \\
	&\barS_{\ell_1 \ell_2}^{\mu \nu,-}[\omega;t]
	= \int \! d\tau \, \theta(-\tau) 
	S_{\ell_1 \ell_2}^{\mu \nu}(t+\tau, t)
	e^{i \int_{t}^{t + \tau} \!\!\! ds\, \omega(s)} \,. 
	\label{eq:omegaRep-}
\end{align}
Although this is certainly 
a generalization of the Fourier components for stationary systems,
infinitely many possible generalizations exist.
However, as discussed in Ref.~\cite{AnnPhys331}, the thermal causality that
the macroscopic time-dependent quantities such as $n_\ell(t)$
affect the microscopic motions in the future only
yields the above expression uniquely.  
As for the index $\ell$, repeating the argument in the equilibrium case that yields the renormalization conditions given in  
Eqs.~(\ref{eq:barS1122onshell}) and (\ref{eq:barS12onshell}), 
we define the
 on-shell self-energy components by setting $\omega(t)=
 \omega_{\ell_1}(t)$ and $\omega_{\ell_2}(t)$ for the retarded part 
$\barS_{\ell_1 \ell_2}^{\mu \nu,+}$ and the advanced part 
$\barS_{\ell_1 \ell_2}^{\mu \nu,-}$, respectively.
First, we impose the renormalization condition
\begin{align}
0&= \barS_{\ell_1 \ell_2}^{11}[\omega_{\ell_1};t]
+\barS_{\ell_1 \ell_2}^{22}[\omega_{\ell_2};t], \notag
\\
&= -2 \delta \omega_{\ell_1 \ell_2}(t) 
+ \barS_{\ell_1 \ell_2}^{11,{\rm loop}}[\omega_{\ell_1};t]
+\barS_{\ell_1 \ell_2}^{22,{\rm loop}}[\omega_{\ell_2};t]
\,,  \label{eq:barS1122noneq}
\end{align}
which fixes $\delta \omega_{\ell_1 \ell_2}(t)$.
As seen for Eq.~(\ref{eq:barS12onshell}),
all the elements of the on-shell 
self-energy with superscript $(1,2)$ vanish as a result of 
the condition given in (\ref{eq:barS1122onshell}) in the equilibrium case, and 
the on-shell self-energy with superscript $(1,2)$
adds no additional restriction. 
We note that $\delta \omega_{\ell_1\ell_2}(t)$ that 
are elements of the Hermitian matrix  generally
have non-zero imaginary parts in off-diagonal
elements $(\ell_1 \neq \ell_2)$.  
While both the completeness and orthonormality of $\{v_{\ell}(x)\}$ 
are retained, the temporal evolution of each $v_{\ell}(x)$ is not represented 
simply by a time-dependent phase factor. In fact, 
the non-vanishing imaginary part of $\delta \omega_{\ell_1\ell_2}(t)$
 $(\ell_1\neq \ell_2)$ plays a crucial role in relaxation of $\{v_{\ell}(x)\}$, which will be confirmed numerically in Section~\ref{sec-model}.
On the other hand,
the quantum transport equation follows from 
the renormalization condition on the 
on-shell self-energy with superscript $(1,2)$. 
Considering the fact that 
the number distribution parameter $n_\ell(t)$
is labeled by a single $\ell$, unlike $\delta \omega_{\ell_1\ell_2}(t)$,
we require application of the renormalization condition 
on the diagonal elements of the on-shell self-energy 
${\bar S}^{12}$ only, such that
\begin{align}
0&=  \barS_{\ell \ell}^{12,+}[\omega_{\ell};t]
+\barS_{\ell \ell}^{12,-}[\omega_{\ell};t], \notag \\
&= -i {\dot n}_\ell(t) + \barS_{\ell \ell}^{12,{\rm loop}}[\omega_{\ell};t]
\,.\label{eq:barS12noneq}
\end{align}
This is the quantum transport equation for a nonequilibrium inhomogeneous
system.

In this section, we summarized the TFD formulation for the nonequilibrium 
inhomogeneous system, using the time-dependent complete orthonormal
set of wave functions. Note that, on deriving the thermal matrix
self-energy according to the Feynman method, we can solve the set of simultaneous equations, (\ref{eq:TDCS}), (\ref{eq:barS1122noneq}), and
 (\ref{eq:barS12noneq}), self-consistently.

\section{Triple-well model with reservoir and results of numerical calculations}\label{sec-model}

Solving the set of simultaneous equations, in particular 
the quantum transport equation (\ref{eq:barS12noneq}), 
requires a large number of numerical calculations, which generates a heavy computational load.
To show that the formulation in the previous section gives
 consistent results and to specifically demonstrate the manner in which the equations are solved,
we consider an open triple-well model coupled with a reservoir, in which 
the positions of the three wells are $x=1\,,\,0\,,-1$\,.
The model Hamiltonian is
\begin{align}
	&H = H_{0} + H_{\rm int} \,, \label{eq:tripleH}\\
	&H_{0}(t) = {\bm \psi}^{\dagger}(t) {\bm h}_{0} {\bm \psi}(t)
	 + \sum_{k=1}^{N} \left(\Omega_{k} - \mu \right)
	 R_{k}^{\dagger}(t) R_{k}(t) \,, \label{eq:tripleH0}\\
	&H_{\rm int}(t) = g \sum_{x = -1}^{1} \sum_{k=1}^{N}
	\left[ R_{k}^{\dagger}(t) \psi_x(t)
	+ \psi_x^{\dagger}(t) R_{k}(t) \right] \,,\label{eq:tripleHint}
\end{align}
where we use column vector notation for the three-component $\psi_x(t)$ 
of the triple-well system
 $(x=1,0,-1)$. Then,
\begin{align}
	{\bm \psi}(t) = \threecolmatrix {\psi_1(t)}{\psi_0(t)}
	{\psi_{-1}(t)}\,, \qquad
	{\bm h}_0= \threematrix {-\mu}{-J}{0}{-J}{-\mu}{-J}
	{0}{-J}{-\mu} \,,
\end{align}
with chemical potential $\mu$ and inter-well hopping $J$.
The operators $R_k$ represent the degrees of freedom of the reservoir.
The canonical commutation relations are 
\begin{align}
	\left[ \psi_{x}(t), \psi_{x'}^{\dagger}(t) \right] = \delta_{xx'}
	\,,\qquad \left[ R_{k}(t), R_{k'}^{\dagger}(t) \right] = \delta_{kk'}
	\,,\qquad {\rm others}=0\,.
\end{align}
In Eqs.~(\ref{eq:tripleH0}) and (\ref{eq:tripleHint}), $\Omega_k$ and $g$ are
the energy spectrum of the reservoir system and the coupling constant
between the triple-well and the reservoir system, respectively. 
The total number of $R_k$, denoted by $N$, is taken
to be $\infty$ at the final calculation stage. At the limit 
$N\,\rightarrow\, \infty$, we replace 
$N\delta_{kk'} \,
 \rightarrow\, \delta(k - k')$ and $1/N \sum_{k=1}^N\, \rightarrow
\, \int_{0}^{k_c} dk$ with $\Omega_{k_c} =\Delta$, 
where $\Delta$ represents the bandwidth of the reservoir energy spectrum.  The coupling constant $g$
 is small and of order $1/\sqrt{N}$, and the finite coupling
constant is defined by
\begin{equation}
	\barg = \sqrt{N} g \,.
\end{equation}
The energy counter term is a $3\times 3$ matrix, and $\dH$ is
\begin{align}
	& \dH ={\bm \psi}^{\dagger}(t) \delta  {\bm \omega}^x(t)
	{\bm \psi}(t) \,,\\
	& \delta {\bm \omega}^x(t)=\threematrix 
	{\delta \omega_{11}(t)}{\delta \omega_{10}(t)}
	{\delta \omega_{1-1}(t)}{\delta \omega_{01}(t)}
	{\delta \omega_{00}(t)}{\delta \omega_{0-1}(t)}
	{\delta \omega_{-11}(t)}{\delta \omega_{-10}(t)}
	{\delta \omega_{-1-1}(t)}\,.
	\label{eq:deltaomegax}
\end{align}
The matrix $\delta {\bm \omega}^x(t)$ is assumed to be Hermitian,
so that $\dH$ is a Hermitian operator.

The field is expanded as
\begin{align}
{\bm \psi}(t)=\sum_{\ell} a_\ell(t)\, {\bm v}_\ell(t) \,.
\end{align}
Here, the wave function ${\bm v}_\ell(t)$, represented by
the column vector
\begin{align}
	{\bm v}_\ell(t) =\threecolmatrix 
	{v_{1\ell}(t)}{v_{0\ell}(t)}{v_{-1\ell}(t)} \,,
\end{align}
is a solution of
\begin{align}
	i\frac {d}{dt} {\bm v}_\ell(t) = 
	{\bm h}_u(t) {\bm v}_\ell(t)-\omega_\ell(t)
	{\bm v}_\ell(t)\, , \label{eq:dotv}
\end{align}
with
\begin{align}
	{\bm h}_u(t)={\bm h}_{0} + \delta {\bm \omega}^x(t)
	\,, \qquad \omega_\ell(t) = {\bm v}^\dagger_\ell(t)
	{\bm h}_u(t){\bm v}_\ell(t)\,.
\end{align}
Note the property ${\bm h}^\dagger_u(t)={\bm h}_u(t)$\,, and that
$\{{\bm v}_\ell(t) \}$ is a complete orthonormal set according 
to the general discussion around 
Eqs.~(\ref{eq:vtOrtht}) and (\ref{eq:vtCompletet}). The three eigenstates are labbeled by $\ell= g,\, o,\, e$, 
as in Appendix A.
Some analytic expressions, necessary for concrete numerical 
calculations but not important for our general formulation, are skipped here
and are given in Appendix.

We move to nonequilibrium TFD, as explained in the previous section,
by doubling every degree of freedom and introducing
the time-dependent number distribution $n_\ell(t)$.
The number distribution of the reservoir $N_{k}$ 
is the Bose-Einstein distribution, such that
\begin{equation}
N_k = \frac{1}{e^{\beta \left(\Omega_k - \mu \right)} - 1 } \,,
\end{equation} 
at temperature $1/\beta$. The thermal counter term ${\hat Q}$ is 
in the unperturbed Hamiltonian, as in Eqs.~(\ref{eq:hatHut}) and
 (\ref{eq:hatQ}).  Following the Feynman diagram method of 
nonequilibrium TFD, we obtain the expression of the
 full propagator of $\psi^\mu_x(t)$, 
$-i \exv{0}{\mrmT \left[ \psi_{\mrmH x_1}^{\mu}(t_1)  
	{\bar \psi}_{\mrmH x_2}^{\nu}(t_2) \right]}$\,, 
from which the corresponding self-energy is extracted, i.e.,
\begin{align}
	&\Sigma_{x_1 x_2}^{\mu\nu}(t_1,t_2) 
	= \left\{-\delta \omega_{x_1x_2}(t_1)\delta_{\mu\nu}
	+i\sum_{\ell} {\dot n}_\ell(t_1) v_{x_1\ell}(t_1) 
	v^\ast_{x_2\ell}(t_2) T_0^{\mu\nu}\right\}
	\delta(t_1-t_2) \notag \\
	&\qquad\qquad\qquad  +\barg^2 \int_{0}^{k_c}
	 dk\, D_k^{\mu\nu}(t_1-t_2) 
	\label{eq:Sigmax1x2}\,,\\
	&T_0= \twomatrix {1}{-1}{1}{-1} \,,
\end{align}
where $D_k^{\mu\nu}$ is the unperturbed propagator of $R_k^\mu(t)$
\begin{align}
	D_{k}^{\mu\nu}(t_1-t_2) 
	=  \left[B^{-1}[N_k] \twomatrix{-i\theta(t_1-t_2)}{0}
	{0} {i\theta(t_2-t_1)} B[N_k]\right]^{\mu\nu} \,
	 e^{-i\Omega_k(t_1-t_2)}\,.
\end{align}
Next, from Eq.~(\ref{eq:Sigmax1x2}), we can derive the self-energy
of $\xi_\ell^\mu(t)$, such that
\begin{align}
	S_{\ell_1\ell_2}^{\mu\nu} (t_1,t_2)
	&=\left\{-\delta \omega_{\ell_1\ell_2}(t_1) 
	{\twomatrix {1}{n_{\ell_2}(t_2)-n_{\ell_1}(t_1)}{0}{1}}^{\mu\nu}
	-i{\dot n}_{\ell_1}(t_1) \delta_{\ell_1\ell_2}
	{\twomatrix {0}{1}{0}{0} }^{\mu\nu}\right\}\delta(t_1-t_2) \notag \\
	& \quad + \barg^2 I^\ast_{\ell_1}(t_1)I_{\ell_2}(t_2)
	\int_{0}^{k_c} dk\, \, e^{-i\Omega_k(t_1-t_2)} \notag \\
	&\qquad \times {\twomatrix {-i\theta(t_1-t_2)}
	{-i\theta(t_1-t_2)n_{\ell_2}(t_2)
	-i\theta(t_2-t_1)n_{\ell_1}(t_1)+iN_k}{0}{i\theta(t_2-t_1)}
	}^{\mu\nu} \,, \label{eq:Sell1ell2}	
\end{align}
where
\begin{align}
	I_{\ell}(t)= \sum_x v_{x\ell}(t) \,.
	\label{eq:Iell}
\end{align}

We substitute (\ref{eq:Sell1ell2}) into 
Eqs.~(\ref{eq:omegaRep}) -- (\ref{eq:omegaRep-}), and for the renormalization conditions of Eqs.~(\ref{eq:barS1122noneq})
and (\ref{eq:barS12noneq}), we explicitly obtain
\begin{align}
	&\domega_{\ell_1 \ell_2}(t) = -i\frac{\barg^2}{2}\int_{0}^{k_c} dk\, 
	\int^t_{-\infty}ds\,\left\{I_{\ell_1}^\ast(t) I_{\ell_2}(s) 
	e^{-i\eta_{k\ell_1}(t,s)}- I_{\ell_1}^\ast(s) I_{\ell_2}(t) 
	e^{-i\eta_{k\ell_2}(s,t)} \right\}\,, \label{eq:NMdomega}\\
	&\dot{n}_{\ell}(t) = -2\barg^2 {\rm Re}\left[ \int_{0}^{k_c}
	 dk\, \ints 
	I_{\ell}^\ast(t) I_{\ell}(s) e^{-i\eta_{k\ell}(t,s)}
	\left\{ n_{\ell}(s) -N_k \right\} \right] \,, \label{eq:NMQTE}\\
	& \eta_{k\ell}(t,s)= \Omega_k(t-s) -\int_s^td\tau\, \omega_\ell(\tau)\,.	\end{align} 
These expressions are non-Markovian in general. We here consider the Markovian limit; 
that is, the system changes so slowly that $I_\ell(s)\,,\, n_\ell(s)$, and 
$\omega_\ell(\tau)$ in the integrands above can be replaced with
$I_\ell(t)\,,\, n_\ell(t)$, and $\omega_\ell(t)$\,. Then, 
Eqs.~(\ref{eq:NMdomega}) and (\ref{eq:NMQTE}) are simplified as
\begin{align}
	&\domega_{\ell_1 \ell_2}(t) = -\frac{\barg^2}{2} I_{\ell_1}^\ast(t)
	 I_{\ell_2}(t) \int_{0}^{k_c}dk\, \biggl[ \mcalP \left\{\frac{1}
	{\Omega_k - \omega_{\ell_1}(t)} + 
	\frac{1}{\Omega_k - \omega_{\ell_2}(t)}\right\}  \notag \\
	& \qquad\qquad\qquad  
	+i\pi \left\{\delta \left( \Omega_k - \omega_{\ell_1}(t)\right) -
	 \delta \left( \Omega_k - \omega_{\ell_2}(t) \right)\right\} \biggr]\,, 	\label{eq:Mdomega} \\
	&\dot{n}_{\ell}(t) =- 2\pi \barg^2 \left| I_{\ell}(t) \right|^2
	\int_{0}^{k_c}dk \, \delta\left( \Omega_k - \omega_{\ell}(t) \right)
	 \left\{  n_{\ell}(t)-N_k \right\}\,. \label{eq:MQTE}
\end{align}
Finally, to proceed to numerical calculations, it is necessary to fix
the $k$-dependence of $\Omega_k$\,. For definiteness, we take the
quadratic form
\begin{equation}
	\Omega_k = k^2  \,.
	\label{eq:Omegak2}
\end{equation}
The substitution of Eq.~(\ref{eq:Omegak2}) into
Eqs.~(\ref{eq:Mdomega}) and (\ref{eq:MQTE}) yields
\begin{align}
	&\domega_{\ell_1 \ell_2}(t) \notag \\
	& \quad 
	= -\frac{\barg^2}{2} I_{\ell_1}^\ast(t) I_{\ell_2}(t)
	\left\{ \bar{C}\left( \omega_{\ell_1}(t) \right) 
	+ \bar{C}\left( \omega_{\ell_2}(t) \right) 
	- i\pi C\left( \omega_{\ell_1}(t) \right) 
	+ i\pi C\left( \omega_{\ell_2}(t) \right) \right\}\,,
	 \label{eq:domegaResult} \\
	&\dot{n}_{\ell}(t) =- 2\pi \barg^2 \left|I_{\ell}(t)\right|^2 
	C\left(\omega_{\ell}(t) \right) 
	\left\{ n_\ell(t) - N (\omega_{\ell}(t))\right\}
	\,,  \label{eq:nResult}
\end{align}
with 
\begin{align}
	C(\omega) = \frac{1}{2 \sqrt{\omega\Delta}}\,, 
	\qquad 	\bar{C}(\omega) =
	 \frac{1}{2\sqrt{\omega\Delta}} 
	\log \frac{ \sqrt{\Delta} - \sqrt{\omega} }
	{\sqrt{\Delta} + \sqrt{\omega}}\,. \label{eq:Comega}
\end{align}

We present the results of numerical calculations 
for the set of simultaneous equations, Eqs.~(\ref{eq:dotv}), 
(\ref{eq:domegaResult}), and (\ref{eq:nResult}).
The latter two equations are derived under the Markovian 
approximation.

We simulate the following nonequilibrium situation: The triple-well 
system and the reservoir are in equilibrium with 
temperature $1/\beta$ for $t<0$, and at $t=0$ the coupling constant 
changes suddenly. For $t>0$, while the reservoir retains the equilibrium
distributions with the same temperature ($1/\beta$), the triple-well system 
follows a nonequilibrium process. We set the parameters as follows: total particle number of triple-well system at initial time 
(and for $t<0$) $\displaystyle \sum_\ell n_\ell=10$;
 $\beta =1/J$; and $\Delta=10 J$.
The coupling constant is primarily fixed to 
$\barg =0.2 J$ for $t<0$ and $\barg=0.1 J$ for $t \geq 0$, but 
$\barg$ for $t<0$ is varied for comparison. 

\begin{figure}[t]
	\centering
	\input{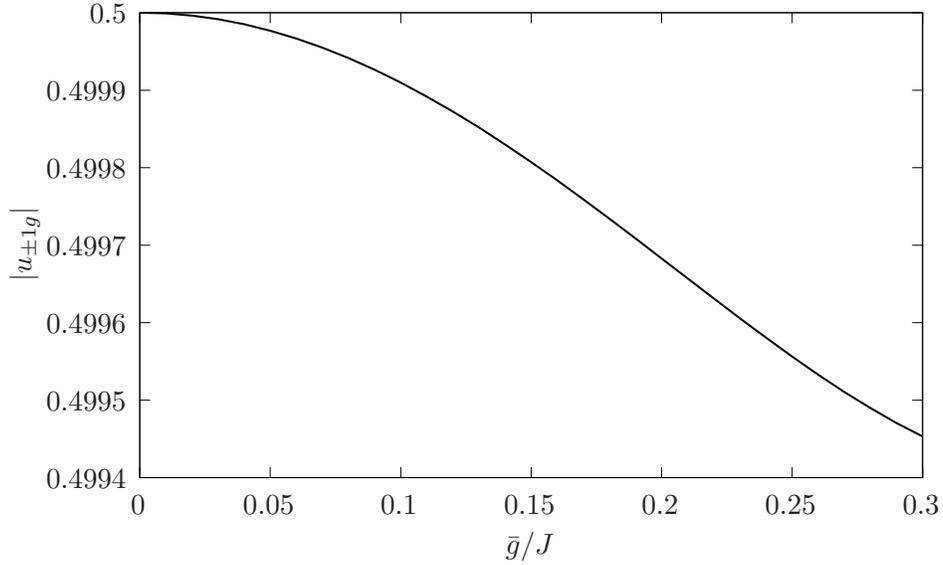}
	\caption{$\barg$-dependence of the components $|u_{\pm 1 g}|$ of the initial
	wave function that is affected by the energy renormalization at equilibrium, 
	This plot shows that $|u_{\pm 1 g}|$ deviates only slightly from 0.5 (the value
	in noninteracting case) for finite ${\bar g}$.}
	\label{fig:initial_CS}
\end{figure}

\begin{figure}[t]
	\centering
	\input{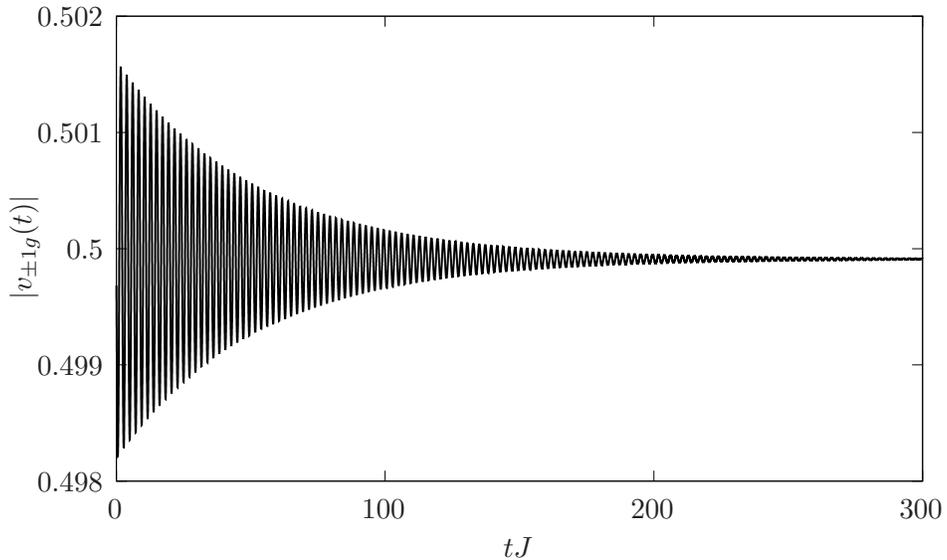}
	\caption{Temporal behavior of $|v_{\pm 1 g}(t)|$}
	\label{fig:TDCS}
\end{figure}

First, we prepare the parameters in the initial equilibrium stage, 
$v_{x\ell}(0)=u_{x \ell}$\,,
\,$\omega_{\ell}(0)$, and $n_{\ell}(0)$, following the 
method given in Section~\ref{sec-InhEqui}. 
The ${\bar g}$-dependence of the eigenfunctions $u_{x\ell}$
that originates solely from the counter term $\delta \omega_{\ell_1\ell_2}$
is weak, as is depicted in Fig.~\ref{fig:initial_CS}.
Then, self-consistent numerical calculations are performed 
to solve Eqs.~(\ref{eq:dotv}), (\ref{eq:domegaResult}), and (\ref{eq:nResult}). 

As regards the results of $\{ v_{x\ell}(t)\}$,
Fig.~\ref{fig:TDCS} shows $|v_{\pm 1g}(t)|$.
The most significant point to note is that 
$|v_{x \ell}(t)|$ relax to the stationary forms of equilibrium. 
The numerical calculations indicate that the non-vanishing imaginary
components of the off-diagonal $\delta \omega_{\ell_1\ell_2}(t)$ 
in Eq.~(\ref{eq:domegaResult}) are crucial for the relaxation of 
$|v_{x \ell}(t)|$. Whereas the counter term 
$\delta \omega_{\ell_1\ell_2}$ affects $v_{x\ell}(0)=u_{x\ell}$ slightly,
as mentioned above, it causes a qualitative change of the temporal behavior
of $v_{x \ell}(t)$ through the imaginary components of its off-diagonal
 elements.
If we renormalized only the diagonal counter 
terms $\delta \omega_{\ell\ell}(t)$, putting 
 $\delta \omega_{\ell_1\ell_2}(t)=0$ $(\ell_1\neq \ell_2)$, the solution of
Eq.~(\ref{eq:dotv}) with the initial condition ${\bm v}_\ell(0)
 ={\bm u}_{\ell}$ would be  a time-dependent phase factor $\times$ 
${\bm u}_\ell$, where
\begin{align}
	{\bm u}_\ell =\threecolmatrix 
	{u_{1\ell}}{u_{0\ell}}{u_{-1\ell}} \,,
\end{align}
and never approaches the final stationary form.

\begin{figure}[t]
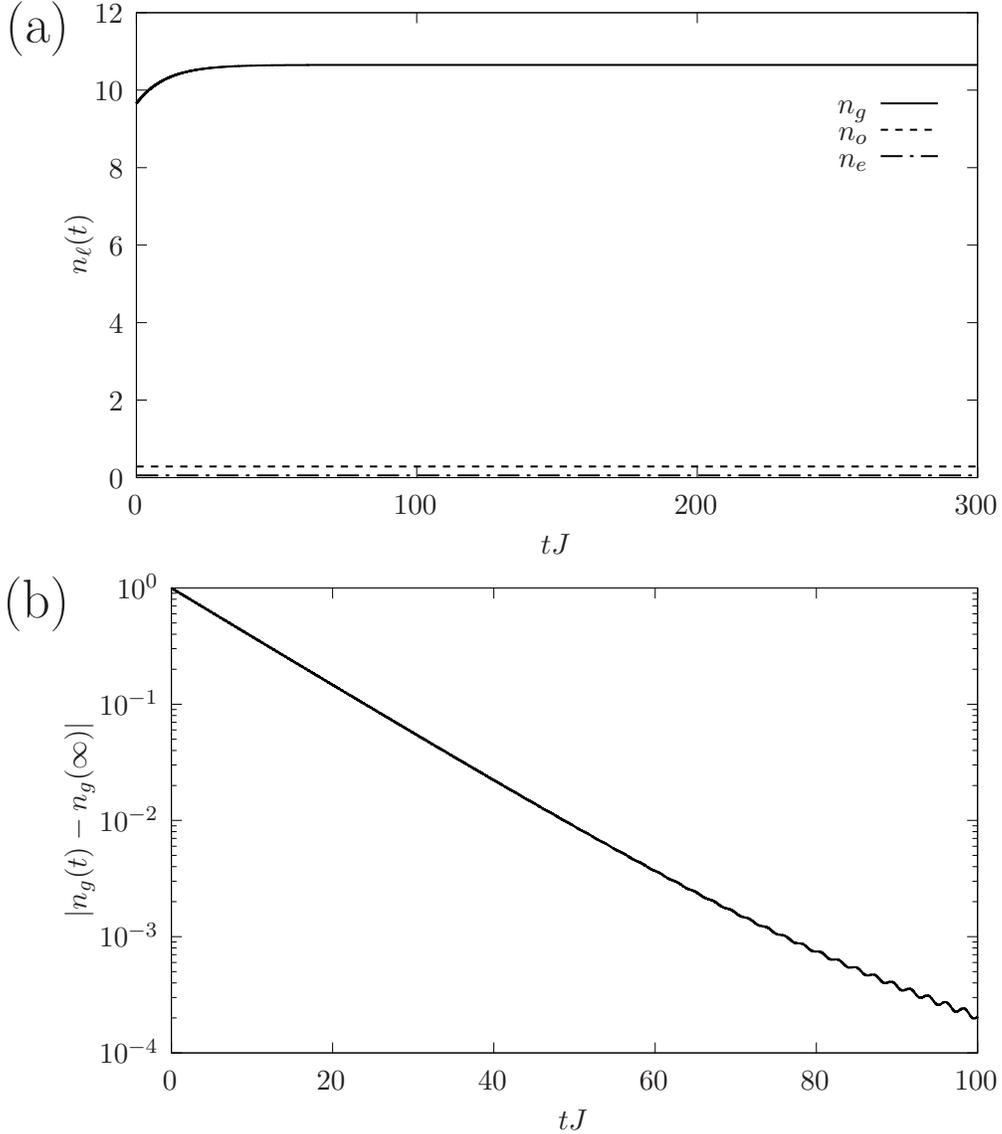

	\centering
	\input{distribution.tex}
	\input{distribution_semi_log.tex}
	\caption{(a) Temporal behaviors of $n_{\ell}(t)$ with $\ell=g,o,e$.
	 (b) Semi-log graph of $|n_{g}(t) - n_{g}(\infty)|$.
}
	\label{fig:distribution}
\end{figure}

Figure~\ref{fig:distribution} (a) shows the variation of 
each number distribution $n_\ell(t)$ $(\ell=g,o,e)$ in time.
Although the changes in $n_o(t)$ and $n_e(t)$ are not clearly visible,
which is due to the selection of a rather low temperature 
and the weak dependence of the eigenfunctions on $\barg$, as shown in 
Fig.~\ref{fig:initial_CS}, $n_g(t)$ approaches a certain value
and the final distributions are the equilibrium distributions with $1/\beta$.
In order to see it clearly, we plot $|n_{g}(t) - n_{g}(\infty)|$ in 
Fig.~\ref{fig:distribution} (b).

It is remarkable that both the number distributions 
and the time-dependent eigenfunctions, following the respective 
equations, are relaxed naturally to the respective equilibrium 
forms at the long-term limit.

\section{Summary}
\label{sec-Summary}

In this paper, we studied the renormalization conditions 
for inhomogeneous systems of a quantum field due to a trapping potential
for both equilibrium and nonequilibrium cases, using the TFD formalism.
Unlike a homogeneous system in which the full propagator, self-energy,
and counter term are diagonal in the momentum index because of the 
total momentum conservation, the inhomogeneous system inevitably 
provides their non-diagonal matrix forms in 
terms of the quantum number $\ell$. Here, it was shown that the $\alpha=1$ representation of TFD 
allows the unique definition of the on-shell self-energy
 in the equilibrium case. Further, the renormalization condition on the on-shell self-energy
with thermal superscripts $(1,1)$ and $(2,2)$ determines 
all the elements of the Hermitian part of the matrix counter term.
Simultaneously, all the elements of the on-shell self-energy with
superscript $(1,2)$ vanish automatically, which is necessary
for extension to nonequilibrium systems.

Next, we discussed the renormalization condition for nonequilibrium
inhomogeneous systems. The core concept was that the nonequilibrium
theory should approach the above equilibrium theory smoothly 
at the long-term limit. The on-shell self-energy in equilibrium
was defined uniquely from a combination of 
the thermal causality and the treatment of the matrix structure
in the equilibrium case. Further, the temporally changing quasiparticle picture naturally generates expansion of the unperturbed field
in terms of the time-dependent wave functions $\{v_\ell(x)\}$. Finally,
we imposed the renormalization condition on the on-shell self-energy
with thermal superscripts $(1,1)$ and $(2,2)$, which specifies all
the matrix elements of the time-dependent energy counter term,
and the other condition on the diagonal part of that 
with thermal superscript $(1,2)$, which gives the quantum 
transport equation for $n_\ell(t)$. Thus, we obtained a set of three coupled 
equations, i.e., the equations for $v_\ell(x)$ and $n_\ell(t)$, and 
the equation to determine the renormalized energy $\omega_\ell(t)$.
It is crucial that the solutions of these equations relax to the 
corresponding equilibrium forms. For the latter, 
the imaginary components of the
 off-diagonal elements of the counter term $\delta \omega_{\ell_1\ell_2}(t)$ 
are crucial.

In order to determine the efficacy of our theory, we performed 
numerical calculations of a triple-well model attached to a reservoir.
The numerical results show that not only $n_\ell(t)$,
but also $v_\ell(x)$ (following their respective equations), approaches
the correct stationary forms at the long-term limit.

There are several research questions related to the formulation
presented in this paper that can be explored in future work. First, we should study more realistic models of cold atomic systems, performing numerical calculations that may have heavy computational loads. In particular, the extension of the present formulation to a cold atomic system with a Bose-Einstein condensate would be very intriguing\cite{PethickSmith,GriffinBook},
because the nonequilibrium phase transition could then be described.
The Bose-Einstein condensation is interpreted as a spontaneous breakdown
of the $U(1)$ global gauge symmetry, and the zero (Nambu-Goldstone) 
mode always appears\cite{UMT,AIP,VitielloBook}. 
As was previously discussed in Ref.~\cite{NTY,AnnPhys376}, the quantum fluctuations of the zero mode cannot be suppressed in a trapped system, but
should be properly incorporated in any analysis. The problem of the renormalization condition including the zero mode remains open.
Further, the extension to relativistic field systems such as
the Klein-Gordon field and Dirac field\cite{PTP126} will find many applications.
\begin{acknowledgments}
This work is supported in part by JSPS KAKENHI Grant No.~16K05488.
The authors thank RIKEN iTHES for offering us the opportunity to discuss 
this work during the workshop on ``Thermal Quantum Field Theories 
and Their Applications" (2016).
\end{acknowledgments}
\appendix

\section{Manupilations of Triple-well model with reservoir}

We summarize some analytic expressions of the model 
in Section~\ref{sec-model}, necessary for  
numerical calculations.

The system symmetry 
under the reflection $x =1\, \leftrightarrow \, -1$ 
restricts the parameters in the
matrix $\delta {\bm \omega}^x(t)$ in 
Eq.~(\ref{eq:deltaomegax}), such that
\begin{align}
&\delta \omega_{11}(t) =\delta \omega_{-1-1}(t)
\,, \qquad
\delta \omega_{1-1}(t) =\delta \omega_{-11}(t), \notag \\
&\delta \omega_{10}(t) =\delta \omega_{-10}(t)
\,, \qquad
\delta \omega_{01}(t) =\delta \omega_{0-1}(t) \,.
\end{align}
Eventually, the matrix $\delta {\bm \omega}^x(t)$ is parameterized
by three real functions $\delta \omega_{11}(t)\,,
\, \delta \omega_{00}(t) \,,\, \delta \omega_{1-1}(t)$
and one complex function $\delta \omega_{10}(t)$, where
\begin{align}
\delta {\bm \omega}^x(t)
=\threematrix 
{\delta \omega_{11}(t)}{\delta \omega_{10}(t)}{\delta \omega_{1-1}(t)}
{\delta \omega^\ast_{10}(t)}{\delta \omega_{00}(t)}{\delta \omega^\ast_{10}(t)}
{\delta \omega_{1-1}(t)}{\delta \omega_{10}(t)}{\delta \omega_{11}(t)}\,.
\label{eq:Symmetrydeltaomega}
\end{align}

We find a constant normalized solution for Eq.~(\ref{eq:dotv}) with 
Eq.~(\ref{eq:Symmetrydeltaomega}), i.e.,
\begin{align}
{\bm v}_o(t) =\frac{1}{\sqrt{2}} \threecolmatrix {1}{0}{-1}\,,
\label{eq:vo}
\end{align}
for it can be confirmed that
$ {\bm h}_u(t) {\bm v}_o= \omega_o(t){\bm v}_o$, with 
$\omega_o(t)=\delta \omega_{11}(t)-\delta \omega_{1-1}(t)$\,,
or that ${\bm h}_0 {\bm v}_o = 0$ and $\delta {\bm \omega}(t) {\bm v}_o
= \omega_o(t){\bm v}_o$\,. This is a single allowed eigenstate with odd parity,
and the remaining two normalized states, having even parities and being orthogonal to
${\bm v}_o(t)$, are generally expressed as
\begin{align}
	&{\bm v}_g(t) = \frac{e^{i \beta_g(t)}}{\sqrt{2(a^2(t)+b^2(t))}}
	\threecolmatrix{ a(t) }{ \sqrt{2} b(t) e^{i \theta(t)}} { a(t) }
	\,, \label{eq:vg}\\
	& {\bm v}_e(t) = \frac{e^{i \beta_e(t)}}{\sqrt{2(a^2(t)+b^2(t))}}
	\threecolmatrix{ b(t) }{- \sqrt{2} a(t) e^{i \theta(t)}} { b(t) }
	\,, \label{eq:ve}
\end{align}
where the five real functions $a(t)\,,\,b(t)\,,\, \theta(t)\,,\,\beta_g(t)$, and $\beta_e(t)$ are determined by solving Eq.~(\ref{eq:dotv}).  
In the stationary (equilibrium) limit, ${\bm v}_g(t)$\,,\,${\bm v}_o(t)\,$, 
and ${\bm v}_e(t)$ become the ground, first excited, and second excited states,
respectively. For later discussions, we introduce
the unitary matrix
\begin{align}
	{\bm V}(t) = \threerowmatrix 
	{{\bm v}_g(t)}{{\bm v}_o(t)}{{\bm v}_e(t)} \,,
	\qquad {\bm V}^\dagger (t) {\bm V}(t) ={\bm V} (t) {\bm V}^\dagger(t)
	={\bm I} \,,
\end{align}
and the Hermitian matrix
\begin{align}
	\delta {\bm \omega}^\ell(t) = {\bm V}^\dagger(t) 
	\delta {\bm \omega}^x(t) {\bm V}(t) 
	=\threematrix {\delta \omega_{gg}(t)}{\delta \omega_{go}(t)}
	{\delta \omega_{ge}(t)}{\delta \omega_{go}^\ast(t)}
	{\delta \omega_{oo}(t)}{\delta \omega_{oe}(t)}
	{\delta \omega_{ge}^\ast(t)}{\delta \omega_{oe}^\ast(t)}
	{\delta \omega_{ee}(t)}\,.
	\label{eq:deltaomegaell}
\end{align}
Each element of this matrix, $\delta \omega_{\ell_1\ell_2}(t)$
($\ell_1,\ell_2=g,o,e$), is determined in the renormalization
condition Eq.~(\ref{eq:barS1122noneq}).  Substituting
 Eqs.~(\ref{eq:Symmetrydeltaomega}), 
(\ref{eq:vg}), and (\ref{eq:ve}) into (\ref{eq:deltaomegaell}),
we obtain 
\begin{align}
&\delta \omega_{go}(t) = \delta \omega_{oe}(t) =0 \,,\\
&\delta \omega_{oo}(t)= 
\delta \omega_{11}(t)-\delta \omega_{1-1}(t)\,.
\end{align}

As regards $I_\ell(t)$ in Eq.~(\ref{eq:Iell}), we note
that the explicit form of ${\bm v}_o(t)$ in Eq.~(\ref{eq:vo})
gives
\begin{equation}
	I_{o}(t)= 0 \,.
\end{equation}

\end{document}